\documentclass[prl,twocolumn,showpacs,amsmath,amssymb,superscriptaddress]{revtex4}
\usepackage{graphicx}
\usepackage{dcolumn}
\usepackage{bm}
\usepackage{color} 

\begin{document}

\title{Topologically trapped vortex molecules in Bose-Einstein condensates}

\author{R. Geurts}
\affiliation{Departement Fysica, Universiteit Antwerpen,
Groenenborgerlaan 171, B-2020 Antwerpen, Belgium}

\author{M. V. Milo\v{s}evi\'{c}}
\affiliation{Departement Fysica, Universiteit Antwerpen,
Groenenborgerlaan 171, B-2020 Antwerpen, Belgium}
\affiliation{Department of Physics, University of Bath, Claverton
Down, Bath, BA2 7AY, UK}

\author{F. M. Peeters}
\email{francois.peeters@ua.ac.be} \affiliation{Departement Fysica,
Universiteit Antwerpen, Groenenborgerlaan 171, B-2020 Antwerpen,
Belgium}

\date{\today}

\begin{abstract}
In a numerical experiment based on Gross-Pitaevskii formalism, we
demonstrate unique topological quantum coherence in optically
trapped Bose-Einstein condensates (BECs). Exploring the fact that
vortices in rotating BEC can be pinned by a geometric arrangement of
laser beams, we show the parameter range in which {\it
vortex-antivortex molecules} or {\it multiquantum vortices} are
formed as a consequence of the optically imposed symmetry. Being
low-energy states, we discuss the conditions for spontaneous
nucleation of these unique molecules and their direct experimental
observation, and provoke the potential use of the phase print of an
antivortex or a multiquantum vortex when realized in unconventional
circumstances.
\end{abstract}


\maketitle

During the last decade, intense experimental and theoretical
activity has induced dramatic achievements in Bose-Einstein
condensation in trapped alkali-metal gases at ultralow temperatures
\cite{bec1}. The atomic Bose-Einstein condensates (BECs) differ
fundamentally from the helium BEC, having significantly non-uniform
density, very high low-temperature condensate fraction, and being
pure and dilute, so that interactions can be accurately parametrized
in terms of the scattering length. As a result, a relatively simple
nonlinear Schr\"{o}dinger equation (the Gross-Pitaevskii (GP)
equation) gives a precise description of the atomic condensates and
their dynamics \cite{LGP}.

It is already well established that the GP equation has rotating
solutions with a well-defined angular momentum, defining the {\it
vortex states} \cite{baym}. Vortices as topological singularities in
BEC have been readily observed in experiment \cite{expBEC}, and are
also found in liquid helium \cite{till} and in superconductors
\cite{tink}. The exciting new property of atomic BECs is that
intrinsic interactions in the system are directly dependent on the
mean particle density \cite{theo}, which gives them a unique degree
of freedom for studies of vortex matter. Another interesting aspect
of dilute Bose condensates is that their density can be locally
suppressed {\it optically}. As experimentally shown by Tung {\it et
al.} \cite{tung}, vortex lines in BEC show great affinity towards
weak spots created by focussed laser beams, which results in a very
profound vortex pinning and provides yet another tool for vortex
manipulation \cite{jap}.

Vortices in BEC are accompanied by the appropriate distribution of
the phase $\varphi$ of the order parameter $\psi$, where the number
of successive $2\pi$ changes along the perimeter corresponds to the
vorticity of the vortex state, i.e. the topological charge.
Therefore, using the known phase distribution, particular vortex
states can be predefined in the system by the so-called {\it phase
imprint} technique \cite{phpr}, where an uniform light pulse
projects a designed mask onto the condensate. However, the gradient
of the imprinted phase ($\nabla\varphi$) must always be parallel to
the rotational velocity of the condensate, i.e. the {\it antivortex}
is fundamentally unstable for a positive angular momentum. Note also
that imprinted multiple $2\pi$ change, i.e. a {\it multiquantum}
vortex, was never found stable against decomposition into single
vortices in conventional BEC setups \cite{butts}.

Motivated by these recent developments in vortex physics, in this
paper we exploit a uniquely defined BEC system that combines above
essentials; our atomic BEC is confined, exposed to engineered
spatial pinning, and is cylindrically rotated. Knowing that vortex
solitons are strongly influenced by symmetry in non-linear media
\cite{ferr}, we impose a particular {\it symmetry} on the system by
chosen arrangement of the vortex-trapping optical beams. As we will
show, this results in the nucleation of vortex molecules of matching
symmetry in the ground state, which may comprise unconventional
phase profiles containing a {\it stable antivortex}, or a {\it
multiquantum} vortex.

For the purpose of this work, we solve the stationary
Gross-Pitaevskii (GP) equation, in a rotating frame with frequency
$\Omega$ around the z-axis:
\begin{equation}
\left( -\frac{\hbar^2}{2m} \Delta + V_{c} + g |\psi|^2 - \Omega L_z
- \mu \right) \psi = 0, \label{gp3d}
\end{equation}
where $V_c$ stands for confinement potential, $g=4 \pi \hbar^2 a /
m$ is the non-linearity with $a$ being the s-wave scattering length,
$L_z=-i\hbar (x\partial_y - y\partial_x)$ is the angular momentum
operator, $\mu$ denotes chemical potential, and $\psi$ is the order
parameter normalized to the number of particles in the condensate
$N$. The energy of a particular state is then given by $E = \mu -
\frac{g}{2} \int dV |\psi|^4 $. We take the usual parabolic-like
confinement potential characterized by the frequency
$\omega_{\perp}$ in the $(x,y)$ plane and $\omega_{z}$ in the
$z$-direction. Our numerical method treats the GP equation in all
three dimensions, but in the present work we take $\omega_z \gg
\omega_{\perp}$ forcing the condensate in a quasi two-dimensional,
oblate (pancake) shape. In that case, Eq. (\ref{gp3d}) can be
written in dimensionless and discretized form as
\begin{widetext}
\begin{eqnarray}
\frac{U_x^{i-1,i}\psi_{i-1,j}}{b_x^2}+\frac{U_x^{i+1,i}\psi_{i+1,j}}{b_x^2}
+\frac{U_y^{j-1,j}\psi_{i,j-1}}{b_y^2}+\frac{U_y^{j+1,j}\psi_{i,j+1}}{b_y^2}
=\left(\frac{2}{b_x^2}+\frac{2}{b_y^2}-g'|\psi_{i,j}|^2-V+\mu\right)\psi_{i,j},
\label{gpdisc}
\end{eqnarray}
\end{widetext}
where $b_{x,y}$ are the lattice constants of the Cartesian grid, and
the link variable is defined as
$U_{\alpha=x,y}^{m,n}=\exp\left[-i\int_{{\bf r}_m}^{{\bf r}_n}{\bf
A}_\alpha({\bf r})d{\bf \alpha}\right]$, with
$A_{x(y)}=(-)\frac{1}{2}\Omega y(x)$. The distances, angular
velocity and energy are expressed in the fundamental scales of the
harmonic trap, i.e. $r_0=\sqrt{\hbar/m\omega_{\perp}}$,
$\omega_0=\omega_{\perp}$, and $E_0=\hbar \omega_{\perp}$
respectively. The 2D nonlinearity $g'=2 a N \sqrt{2\pi
m\omega_{z}\big/\hbar}$ (found by averaging the GP equation over the
z-direction) now depends on $N$ since we chose wave function
normalized to unity. Therefore, for fixed $g$, $N$ is a measure of
the importance of interactions in the condensate, and directly
reflects on the vortex phase diagram. Note also that for low
non-linearities, i.e. when g' is small, vortex states in parabolic
confinement are stable only for velocities of the condensate close
to $\Omega=\omega_{\perp}$, i.e. when the gas dissolves due to
centrifugal forces. To enhance the vortex stability, we have chosen
a modified confinement potential $V_c = \frac{1}{2} r_{\perp}^{2} +
\frac{1}{10}r_{\perp}^{4}$. Such empowered confinement enables the
condensate to survive higher angular velocities
($\Omega>\omega_{\perp}$), and is feasible in experiment by
combining optical and magnetic trapping (e.g. for $^{87}$Rb, a
magnetic trap with frequency $75.5\times 2\pi$ Hz in combination
with a laser beam with waist $25$ $\mu$m and power $1.2$ mW
\cite{squa}).

\paragraph{Vortex-antivortex molecules as the ground state.} In
this paper, we consider $^{23}$Na BEC, with somewhat lower
interaction than $^{87}$Rb, and consequently lower non-linearity.
With the above expression for $V_c$, we use the effective
perpendicular parabolic confinement with $\omega_{\perp}=4.39
\times 2\pi$ Hz, and the unit of length becomes $r_0=10$ $\mu$m.
To obtain the pancake shape of the condensate, we take
$\omega_z=10\omega_{\perp}$. The non-linearity parameter for a
condensate consisting of $10^4$ sodium atoms becomes $g'=44$.
\begin{figure}[b]
\includegraphics[width=\linewidth]{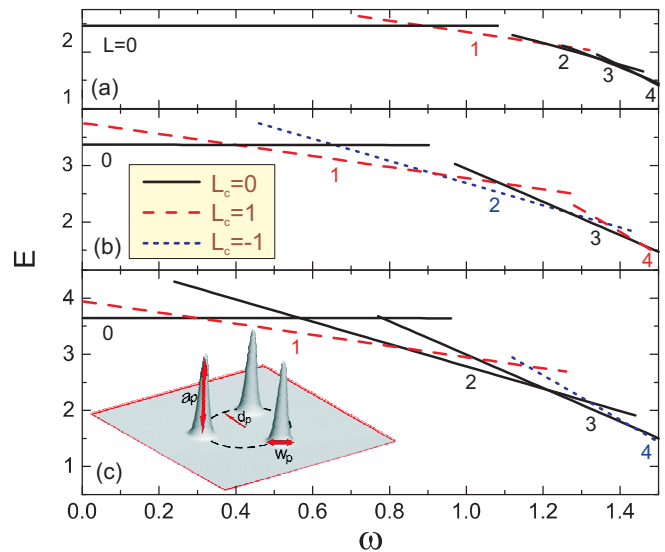}
\caption{\label{fig1} (Color online) Energy versus angular velocity
of a $^{23}$Na condensate with g'=20 (i.e. with $N \approx 5000$
atoms), in the case of no laser pinning ($N_p=0$, a), and three
($N_p=3$, b), and five ($N_p=5$, c) laser beams in a polygonal
setting (with $d_p=0.5$, $w_p=0.4$, $a_p=500$). Different lines
correspond to states with different total vorticity $L$. The colors
and style of the lines indicate the vorticity in the center of the
condensate ($L_c$). The inset depicts the used energy profile for
each pinning site, and notation of the parameters.}
\end{figure}
In what follows, we implement laser pinning in the formalism, and
model it as a Gaussian potential well $V_p = \exp(-16 (r/w_p)^2)$ (r
is the distance from the pinning center) as depicted schematically
in the inset of Fig. \ref{fig1}c. We use a set of $N_p$ laser beams
on a single ring, and compare first the energy in the absence and
presence of pinning, as shown in Fig. \ref{fig1}c. The resulting
total potential in the BEC, introduced by confinement and laser
beams, is taken as
\[ V = V_c(\vec{r}) + a_p \sum_{j=1,N_p} V_p(\vec{r}-\vec{r}_j) \]
with $\vec{r}_j = d_p \left[ \cos{(j2\pi/N_p)} \vec{e}_x +
\sin{(j2\pi/N_p)} \vec{e}_y \right]$. With increasing angular
velocity $\omega = \Omega/\omega_{\perp}$, vortices stabilize in the
system, and each vortex state exhibits a distinct energy. In the
$N_p=0$ case, for chosen parameters, individual vortices form
multi-vortex patterns denoted by $(L_c,L)$, where $L_c$ gives the
number of vortices in the central part of the condensate i.e. $L_c =
\mathop {\lim }\limits_{\epsilon \to 0 } \frac{i}{2 \pi \epsilon}
\mathop{\oint} \limits_{|\vec{r}|=\epsilon} d \left( \log
\frac{\psi}{|\psi|} \right) $, and $L$ is the total vorticity. In
the $N_p\neq 0$ case, the ordering of laser beams imposes its
symmetry on the vortex states, with $N_p$ vortices pinned by the
laser beams, and the remaining ones sitting in the central region of
the condensate. As a remarkable phenomenon, the system may preserve
the symmetry even for $L<N_p$; e.g. for $L=N_p-1$, still $N_p$
vortices nucleate and are pinned, but must be accompanied by a
central {\it antivortex} in order to simultaneously match the total
angular momentum and the symmetry of the pinning potential. Energy
levels of such vortex-antivortex molecules are indicated in Fig.
\ref{fig1} for $N_p=3$ and $N_p=5$ by the dotted (blue) curves.

\begin{figure}[b]
\includegraphics[width=\linewidth]{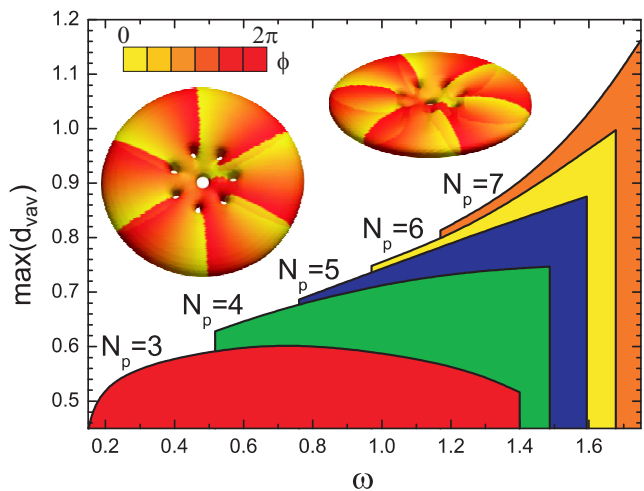}
\caption{\label{fig2} (Color online) Histogram of vortex-antivortex
(V-Av) distance in stable molecules found for different number of
equidistant laser beams on a ring, as a function of angular velocity
for $g'=40$. Maximal V-Av separation increases for more used pinning
beams, and insets show two oblique views of the superimposed
density/phase 3D plots of the condensate for $\omega=1.71$ and
$N_p=7$.}
\end{figure}

Vortex-antivortex (V-Av) phenomena has already been of interest in
BEC, mainly due to the fact that the excitation energy of a
vortex-antivortex pair in non-rotating condensate is {\it lower}
than the one of a single vortex. For that reason, it was assumed
that the actual nucleation of a vortex occurs through the generation
of a V-Av pair and the subsequent expulsion of the antivortex
\cite{vavgen}. While this may hold in ideally non-interacting BEC,
V-Av appearance is seriously hampered by even small perturbations in
the system, particularly in the rotating one. Yet, Crasovan {\it et
al.} demonstrated that a V-Av pair may be stable in an interacting
non-rotating BEC, however {\it only as an excited state} (ground
state remains vortex-free) \cite{vavdipole} and the dynamics of such
a state was recently analyzed in Ref. \cite{vavdipint}. In addition,
vortex-antivortex clusters were studied in Ref. \cite{vavclust}, the
stabilization of vortex-antivortex lattices was considered in Ref.
\cite{vavlattice} and a superposed vortex-antivortex state was
investigated in Ref. \cite{vavsuperpos}, however all only for the
case of {\it non-rotating} systems. Here we show that states
comprising an antivortex can be stabilized as the {\it ground state}
in a rotating BEC by engineered pinning. Moreover, by using a
different number of laser beams, antivortex can be realized {\it at
different angular velocities}. Fig. \ref{fig2} shows the
$\omega$-stability range of V-Av states with one antivortex found
for $N_p=3-7$, thus all states having vorticity $L=N_p-1$. With
increasing $N_p$, not only V-Av states appear for higher angular
momenta, but they can also be spread over a larger area of the
condensate. When artificially spreading the laser beams further
apart (increasing $d_p$, see Fig. \ref{fig1}), vortices
spontaneously follow and increase their distance from the central
antivortex ($d_{vav}$) which facilitates the experimental
verification of the V-Av molecule. Fig. \ref{fig2} also shows the
maximal $d_{vav}$ that can be reached by increasing $d_p$ for $w_p =
0.5$, $a_p=500$ and $g'=40$. The insets of Fig. \ref{fig2} show
isosurfaces of density $10^{-6}$ on which the phase distribution is
superimposed for a condensate in the V-Av state for $N_p=7$, $g'=22$
and $\omega=1.71$.

\begin{figure}[b]
\includegraphics[width=\linewidth]{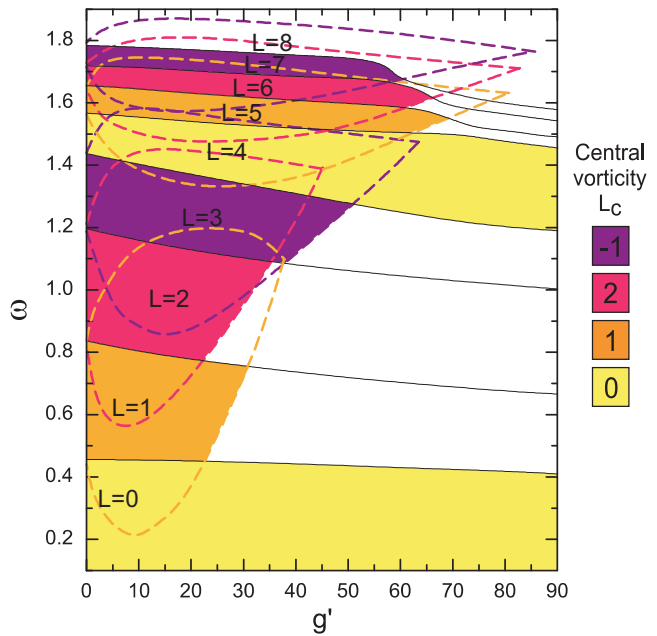}
\caption{\label{fig3} (Color online) Equilibrium vortex phase
diagram as a function of the angular velocity and non-linearity of
the condensate, for $N_p=4$, $d_p=0.71$, $w_p=0.4$ and $a_p=500$.
Coloring scheme illuminates the vorticity in the center of the
condensate, between the pinning beams. Solid lines show the
transitions in the ground state, while dashed curves illustrate the
whole region of stability for different vortex states. In white
areas, the vortex structure does not fully obey the fourfold
symmetry of the pinning.}
\end{figure}

\paragraph{Novel phase transitions.} As much as V-Av states are
novel and exciting study objects, our system actually exhibits even
richer vortex structures beyond V-Av phenomena. To illustrate this,
we constructed a full vortex phase diagram, shown in Fig. \ref{fig3}
as a function of $\omega$ and non-linearity $g'$, for $N_p=4$, i.e.
a square arrangement of laser beams. As shown before, for constant
$g'$, the vorticity increases with increasing $\omega$. Note however
that our strategic pinning setup enforces the $N_p$-fold symmetry of
the vortex states, which leads to specific transitions in the
central region of the condensate. As illustrated by a color gradient
in Fig. \ref{fig3}, the central vorticity changes as $L_c =
\mod(L+1, N_p) - 1$, where an antivortex may nucleate as discussed
above. Nevertheless, $N_p$-fold symmetry tends to remain preserved
for other vorticities as well, which results in the compression of
excess $L-kN_p$ $(k \in \mathbb{N}^{+})$ vortices in the center of
the condensate into a {\it multiquantum vortex} (also called {\it
giant vortex}). This result complements earlier studies of such
vortices in conventional BEC setups, where a multi-quanta vortex was
found {\it unstable} towards decay into single-quantum vortices
\cite{butts}. For $N_p=4$, we show the $0,1,2,-1,0,1,2,-1$ sequence
with increasing $\omega$ in Fig. \ref{fig3}, for total vorticity
$L=0-7$. Doubly quantized vortex was realized for $L=2$ and $L=6$
states in the center of the condensate.

\begin{figure}[b]
\includegraphics[width=\linewidth]{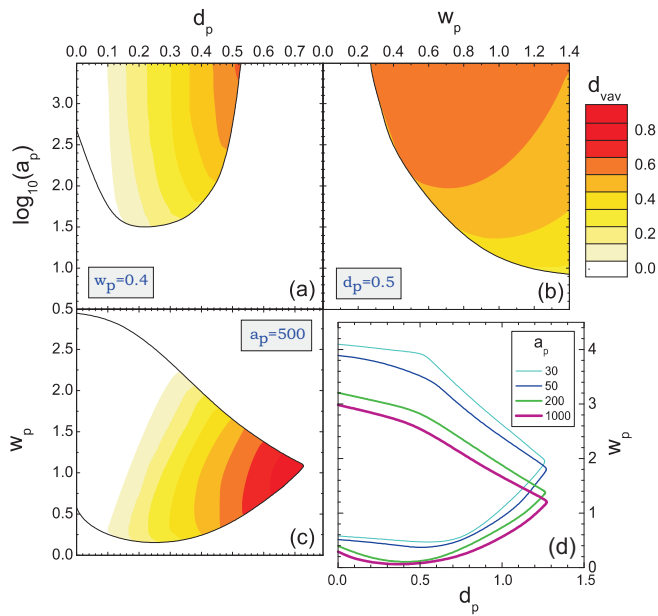}
\caption{\label{fig4} (Color online) Influence of the optical
parameters of the laser beams on the stability of the special vortex
molecules found for $N_p=4$. (a-c) V-Av stability and spacing in a
two-parameter space, thus for either fixed amplitude, separation, or
focused width of the beams ($g'=40$, $\omega=1.25$). (d) The
stability region of the doubly quantized vortex in the center of a
$L=6$ state,
for different values of $a_p$ ($g'=40$, $\omega=1.6$). }
\end{figure}

While the existence of stable multiquantum vortices was already
predicted in several theoretical papers for condensates in a
confinement potential with a quartic component \cite{giantquartic},
our giant vortices are solely a consequence of the symmetry induced
by the pinning centers, as can be deduced from Fig. \ref{fig4}(d)
for the $L=6=4+2$ state: The giant vortex state is not stable when
$w_p$ is too small. By doing two separate simulations of a
condensate in conventional parabolic confinement - thus without
quartic component - exposed to 4 pinning beams, we were able to
confirm this statement: We found a giant vortex in both the ground
states for $L=2$ (with $g = 20$, $\omega=0.85$, $d_p=0.8$,
$w_p=0.6$, $a_p=50$) and $L = 6 = 4 + 2$ ($g = 40$, $\omega=0.85$,
$d_p=1.3$, $w_p=1.6$, $a_p=200$). This general behavior holds for
arbitrary number of pinning beams. In other words, vortex molecules
comprising multiquantum vortices are stable in our system, contrary
to conventional behavior of BEC. Moreover, we report here a unique
way of engineering locally the phase imprint of the condensate, by
either changing angular velocity, or the non-linearity (e.g. by
condensing more constituent atoms). The limits of this procedure
are: (i) the deconfinement of the condensate at large $\omega$, and
(ii) the maximal $g'$ at which vortex-vortex interaction overwhelms
the optical pinning, resulting in a vortex structure that does not
fully obey the imposed geometry (white areas in Fig. \ref{fig3}).

\paragraph{Stability and control of special vortex states.}
Obviously, the exact parameters of the strategically placed laser
beams are very deterministic for the resulting vortex structure in
the condensate. Here we briefly discuss the influence of beam
intensity, distance between adjacent beams, and their focussed
waist on V-Av and multiquantum states. Our results are summoned in
Fig. \ref{fig4}. Figs. (a)-(c) relate to V-Av molecules in
fourfold pinning, and help visualize their stability and size in a
two-parameter space. As a general conclusion, we find that a
larger amplitude of the pinning ($a_p$, see inset of Fig.
\ref{fig1}) is always beneficial to the V-Av state. Simply,
stronger imposed pinning reflects better its symmetry on the
vortex state. For the same reason, strong overlap between pinning
profiles of the beams should be avoided (i.e. $w_p \lesssim 3d_p$
is desirable). As shown in Fig. \ref{fig4}, the distance between
the central antivortex and pinning vortices can be increased by
increasing the spacing between the beams ($d_p$); this is
nevertheless a limited option, as vortices may not follow the
imposed pinning at large distances from the center of the
condensate. For considered parameters, we realized maximal V-Av
distance of approximately $0.9r_0$, which is of the order of
$10\mu$m for the sodium condensate, and thus feasible for
experimental observation. The used beam width of $w_p =
0.2-1.5r_0$ and sub-$r_0$ distances for beam ordering are also
very realistic values for the available experimental techniques.

Fig. \ref{fig4}(d) shows the stability region of a doubly
quantized vortex in a fourfold geometry of the $L=6$ state
($4\times 1+2$). Similarly to the V-Av case, imposed symmetry of
the pinning is essential for the nucleation of the double vortex.
Relatively weaker pinning by wider laser beams can significantly
enhance the stability of the multiquantum vortex, while the
clarity of its separation from the surrounding vortices is best
achieved for high beam amplitudes and narrow beam waists. However,
the needed $d_p$ is quite low, as confinement between the pinning
beams must remain firm for the stabilization of the multiquantum
vortex. Such small spacing is unfavorable for experimental
observation, but can be significantly improved in the case of
larger $N_p$ and larger vorticity. This analysis and the complete
investigation of {\it all} stable vortex states in this
phenomenologically rich system will be presented elsewhere.

\paragraph{Conclusions.}
In summary, using a novel concept of polygonal optical pinning, we
demonstrated yet unpredicted vortex states in rotating atomic
Bose-Einstein condensates. We realized a stable antivortex for
positive angular momentum, as well as a multiquantum vortex, which
up to now was assumed to be unstable in BEC. Both are found as the
ground state for a wide range of parameters. New ground-state
transitions found by increasing angular velocity of the condensate,
by changes in the optical pinning setup, or by changing the number
of constituent atoms, open up ways to further studies employing
quantum phase engineering \cite{phpr} and submicron coherence and
matter-wave interference effects \cite{atomint}.

The authors thank J. Temp\`{e}re for a critical reading of the
manuscript. This work was supported by the Flemish Science
Foundation (FWO-Vl). M.V.M. acknowledges support from the EU
Marie-Curie Intra-European program. The work was partially done
during visit of F.M.P. to the Institute for Mathematical Sciences,
National University of Singapore in 2007.

\end{document}